\documentclass[journal]{IEEEtran}

\usepackage{subfigure}
\usepackage{graphicx}
\usepackage{latexsym}
\usepackage{diagbox}
\usepackage[fleqn]{amsmath}
\usepackage{amsmath}
\usepackage{CJK}
\usepackage{slashbox}
\usepackage{indentfirst}
\usepackage[varg]{txfonts}
\usepackage{stfloats}
\usepackage{multirow}%
\usepackage{booktabs}
\usepackage{color}
\usepackage{color}
\usepackage{epstopdf}
\usepackage{float}
\usepackage{bm}
\usepackage[linesnumbered,ruled,commentsnumbered,longend]{algorithm2e}
\makeatletter

\newcommand{\Rmnum}[1]{\expandafter\@slowromancap\romannumeral #1@}
\makeatother

\ifCLASSINFOpdf
\else
\fi
\begin{document}
%
\title{\huge{Energy-Efficient Power Allocation in  Millimeter Wave Massive MIMO with Non-Orthogonal Multiple Access}}
\author{Wanming Hao, Ming Zeng, Zheng Chu, Shouyi Yang
\thanks{This work was supported by Special Project for Inter-government Collaboration of State  Key Research and Development Program under Grant 2016YFE0118400 and the National Natural Science Foundation of China under Grant U1604159.}
\thanks{W. Hao and S. Yang are with the School of Information Engineering, Zhengzhou University, Zhengzhou 450001, China, and W. Hao is also with the Kyushu University, Fukuoka 819-0395, Japan. (E-mails: wmhao@hotmail.com, iesyyang@zzu.edu.cn)}
\thanks{M. Zeng is with the Faculty of Engineering and Applied Science, Memorial University, St. Johns, NL A1B 3X9, Canada. (E-mail: mzeng@mun.ca)}
\thanks{Z. Chu is with the Faculty of Science and Technology, Middlesex University, London, UK. (Email: z.chu@mdx.ac.uk)}}
\maketitle
\begin{abstract}
In this letter, we investigate the energy efficiency (EE) problem in a millimeter wave (mmWave) massive MIMO (mMIMO) system with non-orthogonal multiple access (NOMA). Multiple two-user clusters are formulated according to their channel correlation and gain difference. Following this, we propose a hybrid analog/digital precoding scheme for the low radio frequency (RF) chains structure at the base station (BS). On this basis, we formulate a power allocation problem aiming to maximize the EE under users' quality of service (QoS) requirements and per-cluster power constraint. {An iterative algorithm is proposed to obtain the optimal power allocation.} Simulation results show that the proposed NOMA achieves superior EE performance than that of conventional OMA.
\end{abstract}


%
\IEEEpeerreviewmaketitle

\section{Introduction}
Power domain non-orthogonal multiple access (NOMA) has been recognized as a promising candidate for next generation wireless communication system~\cite{zeng}. {By applying superposition coding and successive interference cancellation (SIC), NOMA allows multiple users to access the same time-frequency resource, leading to further increase in the spectral efficiency (SE)  compared with orthogonal multiple access (OMA)~\cite{LTE}. Recently, MIMO has been applied to NOMA (MIMO-NOMA) systems to further increase SE \cite{Military, DingmMIMO, Usercluser, EEMIMO}. In a MIMO-NOMA system, users are paired into clusters, with users in each cluster sharing the same beamforming}~\cite{Military, DingmMIMO, Usercluser}.~\cite{Military} proposes a user clustering and power allocation algorithm to maximize the sum capacity. In~\cite{DingmMIMO}, a low-feedback NOMA scheme is proposed for a massive MIMO (mMIMO) system, in which the performance of two scenarios, namely perfect user ordering and one-bit feedback, is evaluated under the proposed scheme.~\cite{Usercluser} jointly investigates user clustering, beamforming design and power allocation problem to maximize the capacity.

However, the above works mainly focus on the SE of the system. With energy efficiency (EE) becoming one of the major concerns for 5G, more efforts need to be paid to its study. So far, only a few works have studied NOMA with the perspective of EE \cite{EEMIMO, EESISO}. In~\cite{EEMIMO}, the authors investigate the EE maximization problem for two users and propose a near optimal power allocation scheme, whereas~\cite{EESISO} considers the EE maximization problem under SISO channel.

Different from previous works, in this letter, we investigate the EE optimization problem in a downlink millimeter wave (mmWave) mMIMO-NOMA system. {Indeed, the use of NOMA in mmWave is prefereable because users' channels can be highly correlated due to the highly directional feature of mmWave transmission~\cite{DingmmWave}. To reduce the hardware complexity, we apply low radio frequency (RF) chains structure at the base station (BS), where the hybrid analog/digitial precoding is considered.} Specifically, we first pair users into clusters according to their channel correlation and gain difference. Following this, we design an analog beamforming vector for each cluster based on the codebook. For digital precoding, {to coordinate inter-cluster interference}, we apply the conventional zero forcing (ZF) precoding based on the channel of the strong user. {To this end, we formulate the power allocation problem aiming to maximize the EE under the users' QoS requirements. To ensure the cluster' fairness,  per-cluster power constraint is considered.} We transform the fractional EE problem into a subtractive-form one, while it is still nonconvex. {Then, we divide the problem into multiple independent convex optimization problems by setting initial feasible power and propose an iterative algorithm to obtain the optimal solution.} Simulation results verify that the EE under NOMA outperforms that under conventional OMA.

\textit{Notation}: We use the following notations throughout this paper:
$(\cdot)^T$ and $(\cdot)^H$ denote the transpose and Hermitian transpose, respectively, $\|\cdot\|$ denotes the Frobenius norm, ${\mathbb{C}}^{x\times y}$ denotes the space of $x\times y$ complex matrix.
\begin{figure}[t]
\begin{center}
\includegraphics[width=8cm,height=3cm]{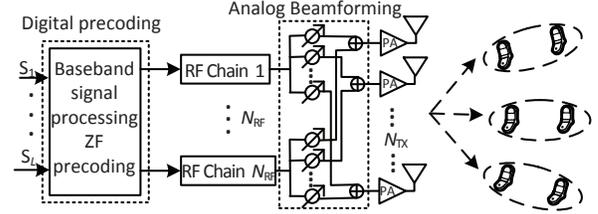}
\caption{Cluster-based downlink mmWave mMIMO-NOMA system model.}
\label{system}
\end{center}
\end{figure}
\section{System and Channel Model}
As shown in Fig. \ref{system}, we consider a downlink mmWave mMIMO-NOMA transmission scenario, in which one BS communicates with $L$ clusters. The BS is equipped $N_{\rm{TX}}$ antennas and $N_{\rm{RF}}$ ($N_{\rm{TX}}\!\geq\!{N_{\rm{RF}}}$) RF chains for reducing the hardware complexity, while each user is equipped with single antenna. The fully connected structure is considered, namely each RF chain is connected to all antennas through a group of phase shifters. We assume that the number of clusters is the same with that of RF chains ($L\!=\!N_{\rm{RF}}$). Following this, users belonging to the same cluster will be supported by the same beamforming vector.
We consider the case that each cluster consists of two users for simplicity, which is also the standard implementation of NOMA in long term evolution advanced (LTE-A)~\cite{LTE}. Specifically, we pair users according to their channel correlation and gain difference~\cite{Military}. {Additionally, we assume that the full channel state information (CSI) is available at BS~\cite{1Gao},~\cite{2Heath}.}

Due to the limited scattering in mmWave channel~\cite{1Gao},~\cite{2Heath}, we adopt a geometric channel model with $F$ scatterers. Each scatter is assumed to contribute a single propagation path between the BS and user. Therefore, the channel ${\bf{h}}_{l,i}~ (i\!\in\!\{1,2\}$, $l\!\in\!\{1,\!\cdots\!,L\})$ can be expressed as:
${\bf{h}}_{l,i}\!=\!\sqrt{{N_{\rm{TX}}}/{F}}\sum_{f=1}^F\beta_{l,i}^f{\bf{a}}(\theta_{l,i}^f),$
where $\beta_{l,l}^f$ denotes the complex gain of the $f$-th path between the BS and the $i$-th user in the $l$-th cluster, which is assumed to be Rayleigh distributed with zero mean and variance of $\sigma_{f}$; $\theta_{l,i}^f\in[0,2\pi]$ is the $f$-th path's azimuth angles of departure (AoDs) of the BS for the user, whereas ${\bf{a}}(\theta_{l,i}^f)$ is the antenna array steering vector with respect to (w.r.t.) $\theta_{l,i}^f$. We only consider the azimuth, but the extension to elevation and azimuth is possible. For an uniform linear array configuration, ${\bf{a}}(\theta_{l,i}^f)$ can  be written as
${\bf{a}}(\theta_{l,i}^f)={{1}/{\sqrt{N_{\rm{TX}}}}}[1,e^{j\frac{2\pi}{\lambda}d\sin(\theta_{l,i}^f)},\ldots, e^{j\frac{2\pi}{\lambda}(N_{\rm{TX}}-1)d\sin(\theta_{l,i}^f)}]^T$,
where $\lambda$ and $d$ denote the signal wavelength and inter-antenna spacing, respectively.

\section{Proposed Hybrid Precoding Design Scheme}
The received signals of the $i$-th user in the $l$-th cluster can be expressed as:
\begin{eqnarray}\label{eq1}
y_{l,i}={\bf{h}}_{l,i}{\bf{B}}{\bf{v}}_{l}s_l+{\bf{h}}_{l,i}\sum_{j\neq l}^L{\bf{B}}{\bf{v}}_{j}s_j+n_{l,i},
\end{eqnarray}
where $s_l\!=\!\sqrt{P_{l,1}}x_{l,1}\!+\!\sqrt{P_{l,2}}x_{l,2}$ denotes the superposed transmitted signals of two users for the $l$-th cluster. $x_{l,1}$, $P_{l,1}$ and $x_{l,2}$, $P_{l,2}$ are the signals and transmit power for User 1 and User 2, respectively. ${\bf{v}}_{l}\in\mathbb{C}^{N_{\rm{RF}}\times 1}$ is the digital precoding vector of the $l$-th cluster, and ${\bf{B}}\in\mathbb{C}^{N_{\rm{TX}}\times N_{\rm{RF}}}$ denotes the analog beamforming matrix for all clusters. $n_{l,i}$ is i.i.d. additive white complex Gaussian noise (AWGN)  with zero mean and variance of $\sigma^2$. {In addition, the second term denotes the inter-cluster interference. Without loss of generality, we assume $\|{\bf{h}}_{l,1}{\bf{B}}{\bf{v}}_{l}\|\geq\|{\bf{h}}_{l,2}{\bf{B}}{\bf{v}}_{l}\|$ in the $l$-th cluster and User 1 and User 2 represent the strong user and weak user,~respectively.}

Different from the conventional analog beamforming design~\cite{2Heath}, where one analog beamforming is designed for one user, two users need to share one analog beamforming in NOMA. In this case, we select a beamforming vector from the codebook of two users that has the best matching with  their overall channels. Since the codebook should have the same form as the array steering vector ${\bf{a}}(\theta_{l,i}^f)$, we define the codebook of the $l$-th cluster as $\mathcal{F}_l=\{{\bf{a}}(\theta_{l,i}^f),f\in\{1,\!\cdots\!,F\},i\in\{1,2\}\}$. On this basis, the analog beamforming of the $l$-th cluster can be selected according to the following criterion:
\begin{eqnarray}\label{eq5}
{\bf{f}}_{{\rm{RF}},l}^\star=\underset{{\bf{f}}_{{\rm{RF}},l}\in\mathcal{F}_l}{\rm{arg\;max}} \;\;\;\|{\bf{h}}_{l,1}{\bf{f}}_{{\rm{RF}},l}\|
+\|{\bf{h}}_{l,2}{\bf{f}}_{{\rm{RF}},l}\|.
\end{eqnarray}

As a result, we can obtain the analog beamforming matrix ${\bf{B}}=[{\bf{f}}_{{\rm{RF}},1}^\star,\cdots,{\bf{f}}_{{\rm{RF}},L}^\star]$.
Then, we can get the equivalent downlink channel of user as $\widetilde{{\bf{h}}}_{l,i}\!=\!{\bf{h}}_{l,i}{\bf{B}}$. Since each cluster consists of two users, the digital precoding can not cancel the inter-cluster interference completely~\cite{Military}. Similar to~\cite{Military}, to perform SIC correctly, we design the digital precoding only considering the channels of the strong users, namely ${\bf{H}}=[\widetilde{{\bf{h}}}_{1,1}^T,\!\cdots\!,\widetilde{{\bf{h}}}_{L,1}^T]^T$. More exactly, we generate the ZF precoding matrix ${\bf{V}}={\bf{H}}^H({\bf{H}}{\bf{H}}^H)^{-1}$, and apply the precoding vector ${\bf{v}}_l={{\bf{V}}(l)}/{\|{\bf{B}}{\bf{V}}(l)\|}$ to the $l$-th cluster, where ${{\bf{V}}(l)}$ denotes the $l$ column of ${{\bf{V}}}$. {As a result, the weak user still receives inter-cluster interference but the strong user does not.}

\section{EE Maximization Problem Formulation And Solution}

{

After hybriding precoding, the received signal-interference-noise-ratio (SINR) of the strong user can be represented as:
${\rm{SINR}}_{l,1}={\|{\bf{h}}_{l,1}{\bf{Bv}}_l\|^2P_{l,1}}/{\sigma^2}.$
In contrast, the weak user receives both inter- and intra-cluster interference, and thus, the received SINR can be written as:
\begin{eqnarray}\label{eq3}
{\rm{SINR}}_{l,2}=\frac{\|{\bf{h}}_{l,2}{\bf{Bv}}_l\|^2P_{l,2}}{\|{\bf{h}}_{l,2}{\bf{Bv}}_l\|^2P_{l,1}+\sum_{j\neq l}\sum_{i=1}^2\|{\bf{h}}_{l,2}{\bf{Bv}}_j\|^2P_{j,i}+\sigma^2},
\end{eqnarray}

{Note that the SIC at the strong user is successful if the strong user received SINR (denoted as ${\rm{SINR}}_{l,2}^1$) for the weak user's signal is larger or equal to the received SINR (denoted as ${\rm{SINR}}_{l,2}^2$) of the weak user for its own signal~\cite{SunUserPair},~\cite{4Heath}. Since ${\rm{SINR}}_{l,2}^1\!=\!{\|{\bf{h}}_{l,1}{\bf{Bv}}_l\|^2P_{l,2}}/({\|{\bf{h}}_{l,1}{\bf{Bv}}_l\|^2P_{l,1}}\!+\!{\sigma^2})$
and ${\rm{SINR}}_{l,2}^2\!=\!{\rm{SINR}}_{l,2}$, it is obvious ${\rm{SINR}}_{l,2}^1\!\geq\!{\rm{SINR}}_{l,2}^2$ according to our assumption $\|{\bf{h}}_{l,1}{\bf{B}}{\bf{v}}_{l}\|\!\geq\!\|{\bf{h}}_{l,2}{\bf{B}}{\bf{v}}_{l}\|$. Thus, the strong user can always remove the weak user's interference successfully.}

Accordingly, the data rates of two users in the $l$-th cluster can be represented as:}
$R_{l,1}=\log_2(1\!+\!\alpha_{l,1}P_{l,1})$, and
$R_{l,2}=\log_2\left(1\!+\!\frac{P_{l,2}}{P_{l,1}\!+\!\sum_{j\neq l}\beta_{l,j}\sum_{i=1}^2P_{j,i}\!+\!1/\alpha_{l,2}}\right)$, respectively,
where $\alpha_{l,i}=\|{\bf{h}}_{l,i}{\bf{Bv}}_l\|^2/{\sigma^2}$, $\beta_{l,j}=\|{\bf{h}}_{l,2}{\bf{Bv}}_j\|^2/\|{\bf{h}}_{l,2}{\bf{Bv}}_l\|^2$.

{The total power consumption consists of two parts: the flexible transmit power and the fixed circuit power consumption.} The fixed part mainly includes baseband, RF chain, phase shifters and power amplifiers (PAs)~\cite{3Heath}, which can be expressed as:
\begin{eqnarray}\label{eq8}
P_{\rm{C}}=P_{BB}+N_{RF}P_{RF}+N_{RF}N_{TX}P_{PS}+N_{TX}P_{PA},
\end{eqnarray}
where $P_{BB}$, $P_{RF}$, $P_{PS}$ and $P_{PA}$ denote the power consumption of the baseband, the RF chain, the phase shifter and {PA}, respectively.
The EE of the system is defined as:
\begin{eqnarray}\label{eq9}
\eta_{\rm{EE}}=\frac{\sum_{l=1}^L(R_{l,1}+R_{l,2})}{\xi\sum_{l=1}^L(P_{l,1}+P_{l,2})+P_{\rm{C}}},
\end{eqnarray}
where $\xi$ is a constant which accounts for the inefficiency of the PA~\cite{xi}.

Finally, we formulate the optimization problem of maximizing EE of the system as follows:
\begin{subequations}\label{eq10}
\begin{align}
&{\rm{max}}\;\;\;\;\eta_{\rm{EE}}\\
{\rm{s.t.}}\;\;&R_{l,i}\geq R_{\rm{min}}, i\in\{1,2\}, l\in\{1,\cdots,L\},\label{eq101}\\
\;\;&P_{l,1}+P_{l,2}\leq P_{\rm{max}}, l\in\{1,\cdots,L\}\label{eq102},
\end{align}
\end{subequations}
where (\ref{eq101}) denote the users' QoS requirements and (\ref{eq102}) are per-cluster power constraints.

We observe that (\ref{eq10}) belongs to a fractional problem, which can be transformed into a parametric subtractive-form problem as:
\begin{eqnarray}\label{eq11}
{\rm{max}}{\sum_{l=1}^L(R_{l,1}\!+\!R_{l,2})}\!-\!\lambda\left({\xi\sum_{l=1}^L(P_{l,1}\!+\!P_{l,2})\!+\!P_{\rm{C}}}\right)\label{eq111},
{\rm{s.t.}}\;(\rm{\ref{eq101}}),(\rm{\ref{eq102}}),
\end{eqnarray}
where $\lambda$ is a non-negative parameter, and we denote the optimal value of (\ref{eq111}) as $T(\lambda)$. With $\lambda^{\rm{opt}}$ denoting the optimal value of (\ref{eq10}), we have the equivalence:
$\lambda=\lambda^{\rm{opt}}\Leftrightarrow T(\lambda)=0,$
which means solving (\ref{eq10}) is equivalent to finding the root for the equation $T(\lambda)=0$. Such equivalence has been proved in~\cite{Nonlinear}, in which an iterative method is proposed to find $\lambda$ by solving the parametric subtractive-form problem at each iteration. Meanwhile, it has been proven that the iterative method converges to the optimal value~\cite{Nonlinear}.

Next, we need to solve the problem (\ref{eq11}) for a given $\lambda$. It is clear that (\ref{eq11}) is a non-convex optimization problem due to the non-concave objective function. We  equivalently  transform  (\ref{eq11}) as follows:
\begin{eqnarray}\label{eq13}
{\rm{max}}\;\;\sum_{l=1}^L\left(R_{l,1}+R_{l,2}-\lambda\xi(P_{l,1}+P_{l,2})\right),\;{\rm{s.t.}}\;(\rm{\ref{eq101}}),(\rm{\ref{eq102}}).
\end{eqnarray}

Since the weak user receives interference from the strong user and other clusters, it is difficult to directly solve the above problem. Thus, we propose an iterative algorithm. We first set the initial feasible solution ${\widehat{\bf{P}}}=[\widehat{P}_{1,1}, \widehat{P}_{1,2}\cdots,\widehat{P}_{L,1},\widehat{P}_{L,2}]$. Following this, the interference from other clusters can be regarded as a constant, (\ref{eq13}) can be transformed into $L$ independent problems as follows:
 \begin{subequations}\label{eq14}
\begin{align}
&\underset{P_{l,1},P_{l,2}}{\rm{max}}\;\;\;\;R_{l,1}+R_{l,2}-\lambda\xi(P_{l,1}+P_{l,2})\label{eq140}\\
{\rm{s.t.}}\;\;&R_{l,i}\geq R_{\rm{min}}, i\in\{1,2\},\label{eq141}\\
\;\;&P_{l,1}+P_{l,2}\leq P_{\rm{max}},\label{eq142}\\
\;\;&P_{j,i}=\widehat{P}_{j,i}, j\neq l, i\in\{1,2\}.\label{eq143}
\end{align}
\end{subequations}

According to the rate expressions of User 1 and User 2, we have:
\begin{eqnarray}\label{eq15}
P_{l,1}\!=\!(2^{R_{l,1}}\!-\!1)/\alpha_{l,1}\; {\rm{and}}\;
P_{l,2}\!=\!(2^{R_{l,2}}\!-\!1)(P_{l,1}\!+\!\Delta \widehat{P}_l\!+\!1/\alpha_{l,2}),
\end{eqnarray}
where $\Delta \widehat{P}_l=\sum_{j\neq l}\beta_{l,j}\sum_{i=1}^2P_{j,i}$.
In the expression of $P_{l,2}$, we substitute $P_{l,1}$ with $(2^{R_{l,1}}-1)/\alpha_{l,1}$ and obtain:
 \begin{eqnarray}\label{eq16}
 P_{l,2}=(2^{R_{l,2}}-1)\left((2^{R_{l,1}}-1)/\alpha_{l,1}+\Delta \widehat{P}_l+1/\alpha_{l,2}\right).
 \end{eqnarray}

Therefore, (\ref{eq14}) can be transformed the following problem:
\begin{subequations}\label{eq17}
\begin{align}
&\underset{R_{l,1},R_{l,2}}{\rm{max}}\;\;\;\;R_{l,1}+R_{l,2}-\lambda\xi f(R_{l,1},R_{l,2})\label{eq170}\\
{\rm{s.t.}}\;\;&({\rm{\ref{eq141}}}), ({\rm{\ref{eq143}}}),\;\;f(R_{l,1},R_{l,2})\leq P_{\rm{max}},\label{eq171}
\end{align}
\end{subequations}
where$f(R_{l,1},R_{l,2})\!=\!P_{l,1}\!+\!P_{l,2}
\!=\!{2^{R_{l,1}\!+\!R_{l,2}}}/{\alpha_{l,1}}\!+\!
[({1}/{\alpha_{l,2}}\!-\!{1}/{\alpha_{l,1}})\!+\!\Delta \widehat{P}_l]2^{R_{l,2}}\!-\!\Delta \widehat{P}_l\!-\!{1}/{\alpha_{l,2}}$.
Since ${1}/{\alpha_{l,2}}\!-\!{1}/{\alpha_{l,1}}\geq 0$, the objective function (\ref{eq170}) is concave w.r.t. $\{R_{l,1},R_{l,2}\}$ and (\ref{eq17}) is a convex optimization problem. {Next, we apply Lagrange dual method~\cite{convex} to solve it. The Lagrangian function of the primal objective function is given by:
\setlength{\mathindent}{0cm}
\begin{align}\label{La}
&L(R_{l,1},R_{l,2},{\bm{\mu}}_l,\theta_l)=R_{l,1}\!+\!R_{l,2}-\lambda\xi f(R_{l,1},R_{l,2})+\mu_{l,1}(R_{l,1}\!-\!R_{\rm{min}})\nonumber\\
&\;\;\;\;\;\;\;\;\;\;\;\;\;\;\;\;\;\;\;+\mu_{l,2}(R_{l,2}\!-\!R_{\rm{min}})+\theta_l(P_{\rm{max}}\!-\!f(R_{l,1},R_{l,2})),
\end{align}
where ${\bm{\mu}}_l\!=\![\mu_{l,1},\mu_{l,2}]$ and $\theta_l$, respectively, are the Lagrange multiplier with minimum date constraints and maximum power constraint in~(\ref{eq171}). Then, the Lagrange dual function can be repressed as
$g({\bm{\mu}}_l,\theta_l)=\underset{\{R_{l,1},R_{l,2}\}}{\rm{max}}L(R_{l,1},R_{l,2},{\bm{\mu}}_l,\theta_l)$
and the dual optimization problem is formulated as:
\begin{eqnarray}\label{dualmin}
\underset{\{{\bm{\mu}}_l,\theta_l\}}{{\rm{min}}}\;\;\;\;g({\bm{\mu}}_l,\theta_l)\;\;\;\;
{\rm{s.t.}}\;\;\;\;{\bm{\mu}}\succeq 0, \theta\geq 0.
\end{eqnarray}

\begin{algorithm}[t]
\caption{{\small{Energy-Efficient Power Allocation Algorithm}}}
\label{algorithms3}
{\bf{Initialize}}  tolerate $\varepsilon$, $\lambda=0$. \\
\Repeat($\left\{\rm{Outer\;iteration}\right\}$){$\varepsilon^\star\leq \varepsilon$}
{Initialize feasible power $\widehat{{\bf{P}}}$.\\
\Repeat($\left\{\rm{Inner\;iteration}\right\}$){$\widehat{{\bf{P}}}$ {\rm{Converge}}}
{Solve (\ref{eq17}) for each cluster by  Lagrange
dual method and obtain ${P_{l,1},P_{l,2}}$, update $\widehat{{\bf{P}}}$.}
Compute $\varepsilon^\star\!=\!\sum_{l=1}^L(R_{l,1}\!+\!R_{l,2})\!-\!\lambda(\xi\sum_{l=1}^L(P_{l,1}\!+\!P_{l,2})\!+\!P_{\rm{C}})$.\\
Update $\lambda=\frac{\sum_{l=1}^L(R_{l,1}+R_{l,2})}{\xi\sum_{l=1}^L(P_{l,1}+P_{l,2})+P_{\rm{C}}}$.\\
}
\end{algorithm}

The sub-gradient-based method~\cite{convex} can be utilized to solve the above dual problem and the updated dual variables in the $s$th iteration can be written as:
\begin{eqnarray}
\begin{aligned}
\mu_{l,i}^{(s+1)}&=\mu_{l,i}^{(s)}+\varphi_{l,i}(s)(R_{l,i}-R_{\rm{min}}), i\in\{1,2\},\\
\theta_l^{(s+1)}&=\theta_l^{(s)}+\phi_l(s)(P_{\rm{max}}-f(R_{l,1},R_{l,2})),
\end{aligned}
\end{eqnarray}
where $\varphi_{l,i}(s)$ and $\phi_l(s)$ are the positive step sizes. For fixed ${\bm{\mu}}_l,\theta_l$, the optimal power allocation can be derived by Karush-Kuhu-Tucker condition~\cite{convex} as follows:
\begin{eqnarray}
P_{l,1}\!=\!\left[\frac{(1\!+\!\mu_{l,1})A_l}{(\mu_{l,2}\!-\!\mu_{l,1})}\!-\!\frac{1}{\alpha_{l,1}}\right]^+,\;
P_{l,2}\!=\!\left[\frac{1\!+\!\mu_{l,2}}{(\xi\lambda\!+\!\theta_l)}
\!-\!\frac{(1\!+\!\mu_{l,2})A_l}{(\mu_{l,2}\!-\!\mu_{l,1})}\right]^+,
\end{eqnarray}
where $A_l\!=\!1/\alpha_{l,2}\!-\!1/\alpha_{l,1}\!+\!\Delta \widehat{P}_l$. Note that we have substituted ${R_{l,1},R_{l,2}}$ into~(\ref{eq15}).

The proposed algorithm includes outer and inner iterations. For inner iteration, since the problem~(\ref{eq17}) is convex, the obtained power is optimal at each iteration. Thus, iteratively updating power will increase or at least maintain the value of the objective function in~(\ref{eq17})~\cite{converge}. Due to the limited transmit power in  each cluster, we will obtain a monotonically non-decreasing sequence with an upper bound (i.e., global optimal) w.r.t. the objective value in~(\ref{eq17}).  On the other hand, the outer iteration always converge to the stationary and optimal solution~\cite{Nonlinear}. Based on the above analysis, the optimal solution can be obtained by the proposed two-layer iterative algorithm, which is summarized as Algorithm 1.

Next, we discuss the complexity of the proposed algorithm. Denote the number of outer and inner iterations as $I_1$ and $I_2$, respectively. For each inner iteration, $L$ subproblems need to be solved through sub-gradient updated-based Lagrange dual method and its complexity is determined by the number of dual variables. Thus, the overall complexity of proposed iterative algorithm is $\mathcal{O}(D^2LI_1I_2)$, where $D$ is the number of dual variables.}

\section{Numerical Results}

Here, we evaluate the EE of the proposed algorithm via simulation. Users are randomly distributed in a cell with a radius of 300 m. The BS is equipped with $N_{\rm{TX}}\!=\!100$ antennas and $N_{\rm{RF}}\!=\!8$ RF chains. We assume that there are $F\!=\!8$ clusters, and the azimuth AOA is uniformly distributed over $[0, 2\pi]$. For the sake of cluster's fairness, the maximum transmit power is set the same for all clusters, and user's QoS is 1 bit/s/Hz. The noise power spectral density is -174 dBm/Hz, and the total available mmWave bandwidth is 50 MHz. In addition, we set $P_{{BB}}\!=\!200$ mW, $P_{{RF}}\!=\!160$ mW, $P_{{PA}}\!=\!40$ mW, $P_{{PS}}\!=\!20$ mW, $\xi$=$1/0.38$, $\sigma_f\!=\!1$ and pass loss exponent 4.3. {As for user pairing, similar to~\cite{Military}, we define the channel correction and gain difference between users $i$ and $j$ as ${\rm{Corr}}_{(i,j)}=|{\bf{h}}_i{\bf{h}}_j^T|/\|{\bf{h}}_i\|\|{\bf{h}}_j\|$ and $\pi_{(i,j)}=\left|\|{\bf{h}}_i\|-\|{\bf{h}}_j\|\right|$, respectively. If ${\rm{Corr}}_{(i,j)}\geq\varepsilon$ ($\varepsilon=0.8$ is the redefined threshold), User $i$ and User $j$ form a cluster. We select $L$ clusters following the decreasing order of ${\rm{Corr}}_{(i,j)}$ (the maximum ${\rm{Corr}}_{(i,j)}$ will be first selected).} We represent conventional OMA with TDMA, where the equal time slots are allocated to users in the same cluster, and the maximum EE is obtained via solving the corresponding convex optimization problem.

\begin{figure}[tb]
  \centering
  \subfigure[]{
    \label{a} 
    \includegraphics[width=4.25cm,height=3.7cm]{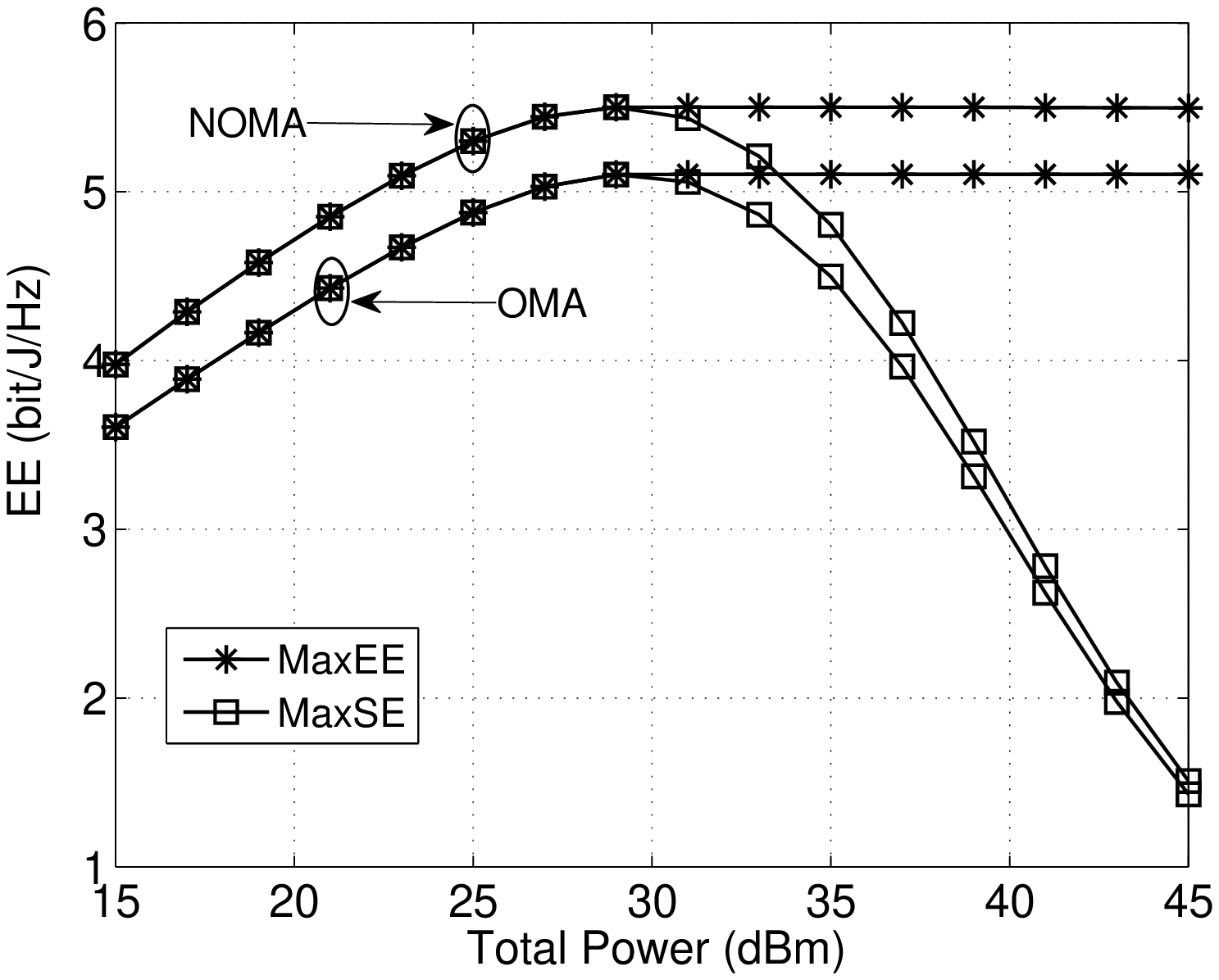}}
  \subfigure[]{
    \label{b} 
    \includegraphics[width=4.25cm,height=3.7cm]{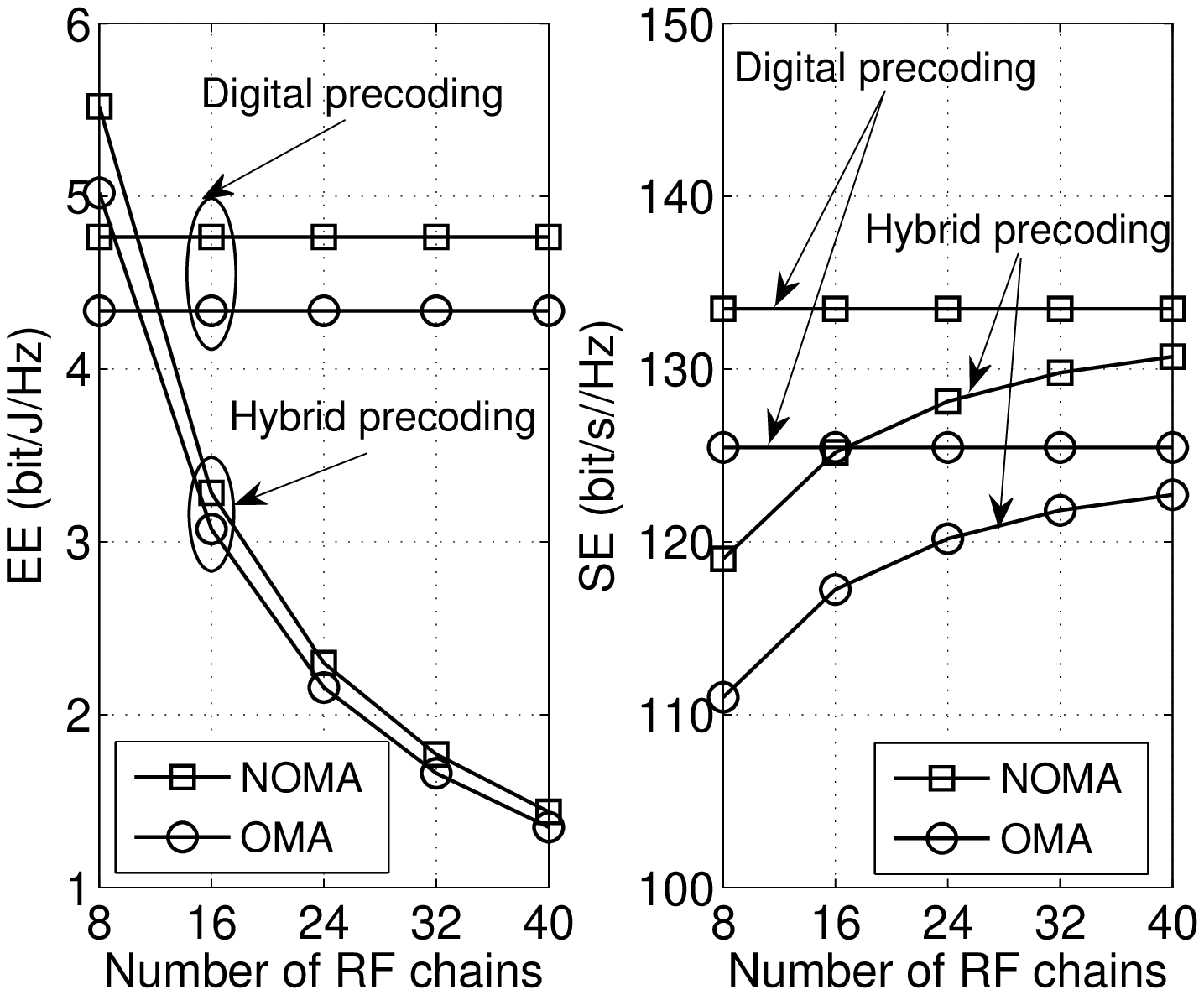}}
  \caption{(a) EE versus total power. (b) EE and SE versus number of RF chains.}
  \label{fig:subfig} 
\end{figure}

Fig.~\ref{a} plots the EE versus the total power. Here, total power denotes the maximum available transmit power of the BS and each cluster is allocated the same power. We assume there are enough users to be selected to form  8 two-user clusters. The legend ``MaxSE" denotes the obtained EE when the SE of the system is maximized, i.e., $\lambda=0$. {It is clear that NOMA outperforms OMA in terms of EE. When the total power is low, increasing it leads to higher EE in both schemes. While, after a certain threshold, the EE reaches a peak and further increase in power brings no enhancement in EE, which indicates that it is not suitable to spend full power from the perspective of EE when the total power is large.
}

{Fig.~\ref{b} shows that the EE versus the number of RF chains when the total power is 30 dBm. Meanwhile, we also present the SE, which is maximized when $\lambda=0$. The legend ``Digital precoding" denotes that each antenna is connected with one RF chain, namely conventional digital precoding structure. Here, we still select 8 two-user clusters, while more RF chains mean that each cluster can obtain more than one analog beamforming. As seen from Fig.~\ref{b}, NOMA achieves higher EE and SE than OMA for both cases. In addition, although more RF chains improve the SE, the EE decreases with the number of RF chains, because more RF chains and phase shifters consume more circuit power.

}

\section{Conclusion}
In this letter, we have investigated the EE maximization problem in a mmWave mMIMO-NOMA system. A hybrid precoding scheme has been designed under the limited RF chains at the BS. On this basis, we have proposed an optimal iterative power allocation algorithm to maximize the EE. Numerical results show that NOMA achieves superior EE performance than OMA, and the EE under hybrid precoding is higher than that under digital precoding for smaller number of RF chains.

\end{document}